\documentclass[11pt,twoside]{article} 
\usepackage{asp2004}
\usepackage{epsf}
\usepackage{psfig}
\usepackage{lscape} 

\begin{document} 
\title{HST and FUSE Spectroscopy of Hot Hydrogen-Rich \\[0.5ex] Central Stars of Planetary Nebulae}
\author{I. Traulsen,$^1$ A.~I.~D. Hoffmann,$^1$ T. Rauch,$^{1,2}$ K. Werner,$^1$ S. Dreizler,$^3$ and J.~W. Kruk$^4$} 
\affil{$^1$Institut f\"ur Astronomie und Astrophysik, Universit\"at T\"ubingen,\\
 Sand 1, D-72076 T\"ubingen, Germany\\
$^2$Dr.-Remeis-Sternwarte, Universit\"at Erlangen-N\"urnberg, Sternwart-\\
str. 7, D-96049 Bamberg, Germany\\
$^3$Institut f\"ur Astrophysik,
    Friedrich-Hund-Platz~1, D-37077~G\"ottingen, Germany\\
$^4$Department of Physics and Astronomy, Johns Hopkins University, Baltimore, MD 21218, USA}

\begin{abstract} 
High-resolution UV spectra, obtained with HST and FUSE, enable us to
analyse hot hydrogen-rich central stars in detail. Up to now, optical
hydrogen and helium lines have been used to derive temperature and
surface gravity. Those lines, however, are rather insensitive; in
particular, neutral helium lines have completely vanished in the
hottest central stars. Therefore, we have concentrated on ionization
balances of metals, which have a rich line spectrum in the UV, to
establish a new temperature scale for our sample. Furthermore, we have
determined abundances of light metals, which had been poorly known
before. They show considerable variation from star to star. We present
results of quantitative spectral analyses performed with non-LTE model
atmospheres.
\end{abstract}

\section{Introduction}

Central stars of planetary nebulae (CSPN) with effective temperatures of more
than 70\,000\,K are in the hottest phase of post-AGB evolution and give
information about late evolutionary stages of sun-like stars, representing the
connective link between Red Giants and White Dwarfs. Caused by the so called
Balmer-line problem and the lack of neutral helium lines in the spectra of the
hottest stars, the results of recent spectroscopic studies on these objects are
sometimes quite uncertain. For these reasons, we have focused on ionization balances of
metals, which show a vast number of lines in the UV, to derive the photospheric
parameters such as $T_\mathrm{eff}$, $\log g$, and light metal abundances of a
sample of seven hot hydrogen-rich CSPN.

\section{Observation and Models}

\begin{figure}
\plotone{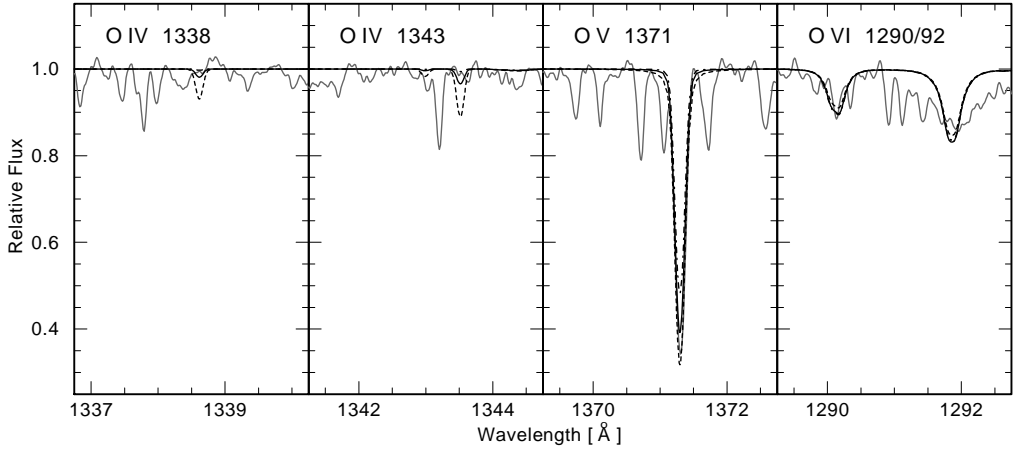}
\caption{\label{teff} Sections of the HST spectrum of NGC\,7293 compared with
  three models with different effective temperatures. The
O\,\textsc{iv}/O\,\textsc{v}/O\,\textsc{vi} ionization balance allows a precise
temperature determination. At $T_\mathrm{eff}=115\,000\,\mathrm{K}$ (dashed),
the computed line cores of the O\,\textsc{iv} triplet at 1338\,/\,1343\,{\AA}
are too deep, whereas the $T_\mathrm{eff}=125\,000\,\mathrm{K}$ model
(dash-dotted) does not fit the O\,\small{v} lines. The best fit is achieved at
$T_\mathrm{eff}=120\,000\,\mathrm{K}$ (full line).}
\end{figure}

With HST and FUSE, we have obtained high-resolution UV and far-UV spectra of
seven hot hydrogen-rich CSPN. Besides, optical spectra of four of the objects
are available (obtained at Siding Spring Observatory, Australia; Hobby-Eberly
Telescope, Texas, USA; Calar Alto Observatory, Spain).

\begin{figure}[b!]
\plotfiddle{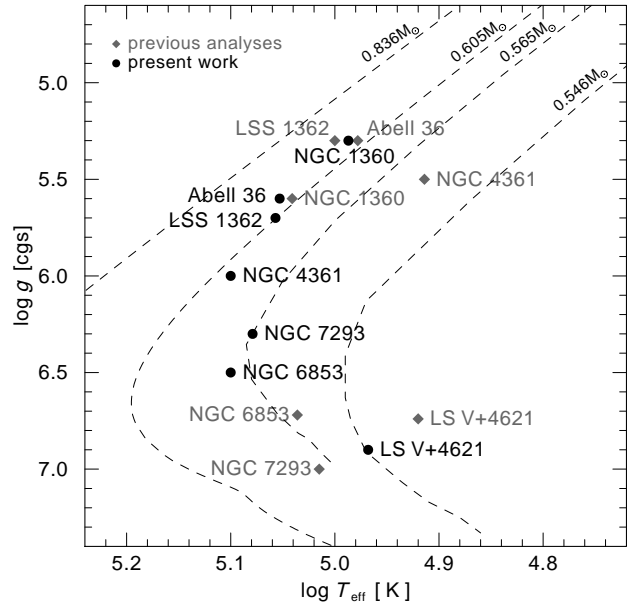}{8.cm}{0.}{40.}{40.}{-120.}{-0.}
\caption{\label{lage} Location of the analysed stars in the $\log g - \log
  T_\mathrm{eff}$ plane. For each object we show the ``old'' location from
  previous analyses and the ``new'' location from the present work. Evolutionary
  tracks are from Sch\"onberner (1983) and Bl\"ocker \& Sch\"onberner (1990).}
\end{figure}

\clearpage

\begin{table}[b!]
\caption{\label{table1} Photospheric parameters of the analysed objects. Due to
  prominent stellar wind features, we suspect that the derived surface gravity
  of LSS\,1362 is quite uncertain. Accuracy of temperature determination is
  5\,\%, of $\log g$ 0.2\,dex, and of element
  abundances better than a factor of 2.\smallskip}
\smallskip
{\small
\begin{tabular}{lrllllll}
\tableline
\noalign{\smallskip}
Star & $T_\mathrm{eff}$ & $\log g$ & $n_\mathrm{He}/n_\mathrm{H}$ & $n_\mathrm{C}/n_\mathrm{H}$ & $n_\mathrm{N}/n_\mathrm{H}$ &
$n_\mathrm{O}/n_\mathrm{H}$ & $n_\mathrm{Si}/n_\mathrm{H}$ \\
 & [K] & [cgs] & & & & & \\
\noalign{\smallskip}
\tableline
\noalign{\smallskip}
LS\,V\,+4621 &  93\,000  & 6.90  & 0.0100 & 0.00004 & 0.000001 & 0.0001  & 0.00002     \\
NGC 1360     &  97\,000  & 5.30  & 0.2500 & 0.0002  & 0.00005  & 0.0002  & 0.00004     \\
Abell 36     & 113\,000  & 5.60  & 0.2000 & 0.0002  & 0.0001   & 0.0005  & 0.000005    \\
LSS 1362     & 114\,000  & 5.70  & 0.1000 & 0.0002  & 0.0001   & 0.0002  & $<$\,0.0004 \\
NGC 7293     & 120\,000  & 6.30  & 0.0300 & 0.0003  & 0.00005  & 0.00035 & 0.000004    \\
NGC 6853     & 126\,000  & 6.50  & 0.1000 & 0.0008  & 0.00002  & 0.0004  & 0.000004    \\
NGC 4361     & 126\,000  & 6.00  & 0.1000 & 0.0080  & 0.00006  & 0.0005  & 0.000002    \\
\noalign{\smallskip}
\tableline
\noalign{\smallskip}
solar  & & & 0.1000 & 0.0004 & 0.0001 & 0.0008 & 0.00004  \\
\noalign{\smallskip}
\tableline
\end{tabular}
}
\end{table}
\begin{table}[b!]
\caption{\label{table2} Additional parameters of the CSPN sample.\smallskip}
\smallskip
{\small
\begin{tabular}{lccccccc}
\tableline
\noalign{\smallskip}
Star & $v_\mathrm{rad}$ & $N_\mathrm{H}$ & $M/M_\odot$ & $R/R_\odot$ & $\log (L/L_\odot)$ & $d$ &
$M_\mathrm{V}$ \\
 & [km/s] & [cm$^{-2}$] & & & & [kpc] & [mag] \\
\noalign{\smallskip}
\tableline
\noalign{\smallskip}
LS\,V\,+4621 & $+22.4$ & $8.6\cdot 10^{19}$ & 0.55 & 0.04 & 2.10 & 0.24 & 5.9 \\
NGC 1360    & $+48.1$ & $5.1\cdot 10^{19}$ & 0.65 & 0.30 & 3.64 & 0.93 & 1.8 \\
Abell 36    & $+36.5$ & $4.9\cdot 10^{20}$ & 0.65 & 0.21 & 3.70 & 0.77 & 2.5 \\
LSS 1362    &  $-2.7$ & $4.2\cdot 10^{20}$ & 0.60 & 0.18 & 3.70 & 1.03 & 2.6 \\
NGC 7293    &  $-5.5$ & $1.1\cdot 10^{20}$ & 0.56 & 0.09 & 3.13 & 0.78 & 4.3 \\
NGC 6853    & $-23.1$ & $1.3\cdot 10^{20}$ & 0.58 & 0.07 & 3.20 & 0.94 & 4.5 \\
NGC 4361    & $+16.4$ & $1.2\cdot 10^{20}$ & 0.59 & 0.13 & 3.65 & 1.08 & 3.2 \\
\noalign{\smallskip}
\tableline
\end{tabular}
}
\end{table}

Spectral analyses are performed with plane-parallel, line-blanketed
non-LTE model atmospheres in hydrostatic and radiative equilibrium,
calculated by {\small NGRT} and other programs of the T\"ubingen NLTE
Model Atmosphere Package (Werner \& Dreizler 1999). Due to
deficiencies in line-broadening theory and in the knowledge of levels and
transitions of highly ionized elements in the UV, atomic data
particularly of O\,\textsc{vi} is uncertain. In comparison with the
high-resolution spectra, we have been able to revise and adjust the
predicted wavelengths of several O\,\textsc{vi} lines.

\section{Results}

Ionization balances of light metals, in particular those of
O\,\textsc{iv}/O\,\textsc{v}/O\,{\textsc{vi}
(Fig.\nolinebreak\,\ref{teff}), enabled us to determine the
photospheric parameters of our sample rather precisely and to
establish a new temperature scale. Besides the surface gravity, we
derived element abundances of helium, carbon, nitrogen, oxygen, and
silicon. A wide spread of abundances from 0.05 solar up to 20 times
solar shows the variety in chemical compositions of hot hydrogen-rich
CSPN.

Most of the objects show considerably higher effective temperatures
than previously thought (Fig.\,\ref{lage}). We determined temperatures
from 90\,000\,K up \linebreak to almost 130\,000\,K. Surface gravity
covers the range between $\log g=5$ and $\log g=7$ [cgs]. In
comparison with evolutionary theory, we have furthermore derived
stellar parameters such as mass, radius, and luminosity. The results
are shown in Tables \ref{table1} and \ref{table2}.

Details about these analyses can be found in Traulsen (2004).
We are also modeling the spectral lines from iron group elements. From these we
will determine heavy metal abundances and we can further improve the
temperature scale of the CSPN sample (Hoffmann et\,al., these proceedings).

\begin{figure}[t!]
\plotone{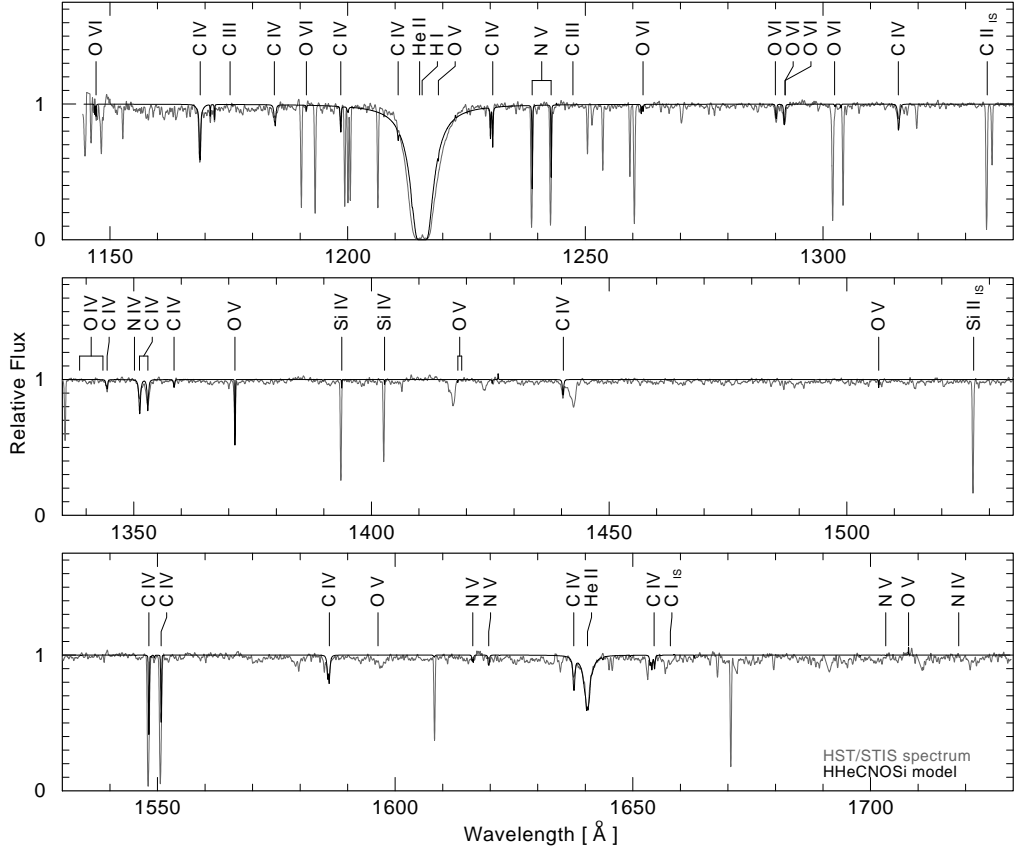}
\caption{\label{ngc4361} NGC\,4361 has the highest temperature in our 
sample. It appears to be a metal-poor halo object with a remarkable
overabundance 
in carbon.}
\end{figure}

\acknowledgements{TR is supported by the DLR (grant 50\,OR\,0201) and
JWK by the FUSE project, funded by NASA contract NAS5-32985.}

\end{document}